\documentclass[aps,amsmath,floatfix,twocolumn,showpacs]{revtex4}

\usepackage{graphicx}
\usepackage{euscript}
\usepackage{oldgerm}
\usepackage{subfigure}
\usepackage{color}
\usepackage{txfonts}
\usepackage{amssymb}

\begin{document}

\title{Cooperation of Sperm in Two Dimensions: Synchronization,
Attraction and Aggregation through Hydrodynamic Interactions}
\date{\today}
\author{Yingzi Yang, Jens Elgeti}
\author{Gerhard Gompper} \email[]{e-mail: g.gompper@fz-juelich.de}
\affiliation{Theoretical Soft Matter and Biophysics Group, Institut
f{\"u}r Festk{\"o}rperforschung, Forschungszentrum J{\"u}lich,
D52425 J{\"u}lich, Germany}

\begin{abstract}
Sperm swimming at low Reynolds number have strong hydrodynamic
interactions when their concentration is high in vivo or near
substrates in vitro. The beating tails not only propel the sperm
through a fluid, but also create flow fields through which sperm
interact with each other. We study the hydrodynamic interaction and
cooperation of sperm embedded in a two-dimensional fluid by using a
particle-based mesoscopic simulation method, multi-particle
collision dynamics (MPC). We analyze the sperm behavior by
investigating the relationship between the beating-phase difference
and the relative sperm position, as well as the energy consumption.
Two effects of hydrodynamic interaction are found, synchronization
and attraction. With these hydrodynamic effects, a multi-sperm
system shows swarm behavior with a power-law dependence of the
average cluster size on the width of the distribution of beating
frequencies.
\end{abstract}


\maketitle

\section{Introduction}
Sperm motility is important for the reproduction of animals. A
healthy mature sperm of a higher animal species usually has a
flagellar tail, which beats in a roughly sinusoidal pattern and
generates forces that drive fluid motion. At the same time, the
dynamic shape of the elastic flagellum is influenced by the fluid
dynamics. The snake-like motion of the tail propels the sperm
through a fluid medium very efficiently. In the past decades, the
effort to quantitatively describe the fluid dynamics of sperm has
been very successful \cite{GRAY1955,FauciRev}.

However, despite considerable progress in modeling sperm elementary
structures and the behavior of a single sperm in a fluid medium
\cite{Brokaw2001,Camalet1999}, relatively few studies have examined
the fluid-dynamics coupling of sperm and other mesoscopic or
macroscopic objects, e.g., the synchrony of beating tails
\cite{Taylor,Fauci}, the tendency of accumulation near substrates
\cite{rothschild63,Woolley2003,Jens,Berke}, etc. In nature, the
local density of sperm is sometimes extremely high. For example, in
mammalian reproduction, the average number of sperm per ejaculate is
tens to hundreds of millions, so that the average distance between
sperm is on the scale of ten micrometers --- comparable to the
length of their flagellum. The sperm are so close that the
interaction between them is not negligible. In recent years,
experiments \cite{Hayashi,Moore,Hayashi2,Immler,Moore2,Riedel} have
revealed an interesting  swarm behavior of sperm at high
concentration, e.g. the distinctive aggregations or 'trains' of
hundreds of wood-mouse sperm \cite{Immler,Moore2}, or the vortex
arrays of swimming sea urchin sperms on a substrate \cite{Riedel}.
The mechanisms behind the abundant experimental phenomena are still
unclear. In this paper, we focus on the hydrodynamic interaction
between sperm and explain its importance for the cooperative
behavior.

The higher animal sperm typically have tails with a length of
several tens of micrometers. At this length scale, viscous forces
dominate over inertial forces. Thus, the swimming motion of a sperm
corresponds to the regime of low Reynolds number \cite{Purcell}.
Experimental observations of two paramecium cells swimming at low
Reynolds number have shown that the changes in direction of motion between
two cells are induced mainly by hydrodynamic forces \cite{Ishikawa}.
Studies of model micro-machines indicate that hydrodynamic
interaction is significant when the separation distance
is comparable to their typical size \cite{Nasseri}.
The hydrodynamic interaction between two rotating helices,
like bacterial flagella, has been investigated both experimentally
\cite{Kim2} and theoretically \cite{Kim,Reichert05}.
An artificial microswimmer, which mimics the motion of a beating sperm,
has been constructed from a red blood cell as head and a flagellum-like tail
composed of chemically linked paramagnetic beads;
the propulsion is then induced by a magnetically driven undulation
of the tail \cite{Dreyfus05}. Simulations have been employed to
study the motion of a single of these artificial microswimmer
\cite{Gauger06}, as well
as the hydrodynamic interactions between two swimmers \cite{Keaveny08}.
Even studies of a minimal swimming model of three linearly connected
spheres \cite{Najafi} have shown a
complicated cooperative behavior \cite{Pooley}. Thus, although there
has been much progress on modeling and observing a single swimmer,
the understanding on the hydrodynamic coupling behavior of dense
system of swimmers is still poor.

In this paper, we focus on the cooperation behavior of sperm in two
dimensions. Although real swimming spermatozoa are certainly
three-dimensional, the qualitatively similar phenomena, and the
great saving of simulation time, makes it worthwhile to discuss the
problem of cooperation in a viscous fluid in two dimensions.
Furthermore, sperm are attracted to substrates in in-vitro experiments
\cite{rothschild63,Woolley2003,Jens,Berke}
and are therefore often swimming under quasi-two-dimensional condition
(it has to be emphasized that hydrodynamic interactions in two dimensions
and in three dimensions near a substrate are of course different).
Thus, we construct a
coarse-grained sperm model in two dimensions and describe the motion
of the surrounding fluid by using a particle-based mesoscopic
simulation method called multi-particle collision dynamics (MPC)
\cite{Malevanets,Malevanets2}.  This simulation method has been
shown to capture the hydrodynamics and flow behavior of complex
fluids over a wide range of Reynolds numbers very well
\cite{Lamura,Ripoll}, and is thus very suitable for the simulation
of swimming sperm.

This paper is organized as follows. Section~II gives a brief description
of our sperm model and of the particle-based hydrodynamics approach.
In order to understand a complex many-body system of micro-swimmers, a
first-but-important step is to investigate the interaction between
two swimming sperm.  Thus, in Sec.~III,
we look at the cooperative behavior of two sperm.  Two remarkable
hydrodynamic effects, synchronization and attraction, are found and
discussed in detail.  In Sec.~IV, we analyze the clustering behavior
of multi-sperm systems. In particular, we consider a sperm system
with a distribution of beating frequencies, and determine the
dependence of the cluster size on the variance of the frequency
distribution.

\section{Sperm Model and Mesoscale Hydrodynamics}\label{model}

\subsection{Multi-Particle Collision Dynamics (MPC)}

MPC is a particle-based mesoscopic simulation technique to describe the
complex fluid behaviors for a wide range of Reynolds numbers
\cite{Lamura,Ripoll}. The fluid is modeled by $N$ point particles, which
are characterized by their mass $m_\text{i}$,
continuous space position ${\bf r}_\textit{i}$ and continuous
velocity ${\bf u}_\textit{i}$, where \textit{i} = 1,\ldots $N$. In
MPC simulations, time $t$ is discrete. During every time step
$\Delta$\textit{t}, there are two simulation steps, streaming and
collision. In the streaming step, the particles do not interact with
each other, and move ballistically according to their velocities,
\begin{equation}
   {\bf r}_{i}(t+\Delta t)={\bf r}_{i}(t)+{\bf u}_{i}\Delta t
\end{equation}
In the collision step, the particles are sorted into collision boxes
of side length $a$ according to their position, and interact with
all other particles in same box through a multi-body collision. The
collision step is defined by a rotation of all particle velocities
in a box in a co-moving frame with its center of mass. Thus, the
velocity of the $i$-th particle in the $j$-th box after collision
is
\begin{equation}
   {\bf u}_{i}(t+\Delta t)={\bf u}_{cm,j}(t)+
           \mathfrak R_j(\alpha)[{\bf u}_{i}-{\bf u}_{cm,j}]
\end{equation}
where
\begin{equation}
   {\bf u}_{cm,j}(t)=\frac{\sum_{j} m_{i}{\bf u}_{i}}{\sum_{j} m_{i}}
\end{equation}
is the center-of-mass velocity of $j$-th box, and
$\mathfrak R_j(\alpha)$ is a rotation matrix which rotates a vector
by an angle $\pm\alpha$, with the sign chosen randomly. This implies
that during the collision each particle changes the magnitude and
direction of its velocity, but the total momentum and kinetic
energy are conserved within every collision box. In order to ensure
Galilean invariance, a random shift of the collision grid has to be
performed \cite{Ihle,Ihle2}.

The total kinematic viscosity $\nu$ is the sum of two contributions,
the kinetic viscosity $\nu_{kin}$ and the collision viscosity
$\nu_\textit{coll}$. In two dimension, approximate analytical
expressions are \cite{Kikuchi,Tuezel},
\begin{eqnarray}
  \frac{\nu_{coll}}{\sqrt{k_{B}Ta^2/m}} &=&
    \frac{1}{12 h} \, {(1-\cos{\alpha})}\, \biggl(1-\frac{1}{\rho}\biggr) \\
  \frac{\nu_{kin}}{\sqrt{k_{B}Ta^2/m}} &=&
    h \, \biggl[\frac{1}{1-\cos\alpha}\frac{\rho}{\rho-1}-\frac{1}{2}\biggr]
\end{eqnarray}
where $\rho$ is the average particle number in each box, $m$ is the
mass of solvent particle and $h=\Delta t\sqrt{k_BT/ma^2}$ is the
rescaled mean free path. In this paper, we use $k_BT=1$, $m=1$,
$a=1$, $\Delta t=0.025$, $\alpha=\pi/2$, $\rho=10$.
This implies, in particular, that the simulation time unit
$(ma^2/k_BT)^{1/2}$ equals unity. With these parameters, the total
kinematic viscosity of fluid is $\nu=\nu_{coll}+\nu_{kin}\approx3.02$.
The size of the simulation box is $L_x\times L_y$, where $L_x=L_y=200a$,
four times the length of the sperm tail, if not indicated otherwise.
Periodic boundary conditions are employed.

\subsection{Sperm Model in Two Dimensions}

\begin{figure}
\includegraphics[width=6cm]{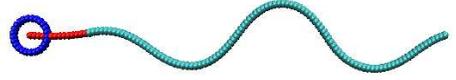}
\caption{(Color online) Two-dimensional model of sperm. The model
consists of three parts, the head (blue), the mid-piece (red) and
the tail (cyan). Two sinusoidal waves are present on the beating
tail.} \label{spermmodel}
\end{figure}

Although animal sperm differ from species to species,
their basic structure is quite universal. Usually, a sperm consists
of three parts: a head containing the genetic information, a beating
long tail, and a mid-piece to connect head and tail.
Our two-dimensional sperm model, shown in Fig.~\ref{spermmodel},
consists of these three parts. The head is constructed of
$N_{head}=25$ particles, where neighboring particles are linked by
springs of finite length $l_0=0.5a$ with interaction potential
\begin{equation}
V_{bond}({\bf R}) = \frac{1}{2} k (|{\bf R}| - l_0)^2
\end{equation}
into a circle of radius
$2a$. Each of the head particles has a mass $m_{head}=20$. The
mid-piece consists of $N_{mid}=14$ particles of mass $m_{mid}=10$
connected by springs of length $l_0=0.5a$. The first
particle of the mid-piece, which is fixed to the center of the head,
is connected with every particle on the head by a spring of length
$l_{head-mid}=2a$, in order to maintain the circular shape of the
head, as well as to stabilize the connection between head and mid
part. The tail has $N_{tail}=100$ particles of mass $m_{tail}=10$,
linked together by springs of length $l_0$.
The spring constants are chosen to be $k_{head-mid}=10^4,
k_{head}=10^5, k_{mid}=k_{tail}=2\times10^5$, where $k_{head-mid}$
is the spring constant for the connection of the head particles and
the center, and $k_{mid}$ and $k_{head}$ are the spring constants for
the tail and the mid-piece, respectively.

A bending elasticity is necessary for the mid-piece and the tail to
maintain a smooth shape in a fluctuating environment, and to implement
the beating pattern. The bending energy is
\begin{eqnarray}
E_{bend} &=& \sum_{i\,\in\, mid}\frac {1}{2}\kappa
                 \biggl \{ {\bf R}_{i+1}- {\bf R}_i\biggr \}^2\\
 & & +\sum_{i\,\in\, tail}\frac {1}{2}\kappa
      \biggl \{ {\bf R}_{i+1}-\mathfrak R(l_0c_{s,tail}) {\bf R}_i\biggr \}^2
\end{eqnarray}
where $\kappa$ denotes the bending rigidity, $\mathfrak R(l_0c_{s})$
is a rotation matrix
which rotates a vector anticlockwise by an angle $l_0c_{s}$, and
$c_{s,tail}$ is the local spontaneous curvature of the tail of the
$s$-th sperm.
We choose $\kappa = 10^4$, much larger than the thermal energy
$k_BT=1$ to guarantee that the mechanical forces dominate the
thermal forces.
For the mid part, the spontaneous curvature vanishes,
$c_{s,mid}=0$. $c_{s,tail}$ is a variable changing with time $t$ and the
position $x$ along the flagellum to create a propagating bending wave,
\begin{align}
    c_{s,tail}(x,t)=c_{0,tail}+A\sin\biggl[-2\pi f_s t+qx+\varphi_s\biggr] .
\end{align}
A detailed analysis of the beating pattern of bull sperm in
Ref.~\cite{Riedel-Kruse07} shows that a single sine mode represents
the beat to a very good approximation. The wave number
$q=4\pi/l_0N_{tail}$ is chosen to mimic the tail shape of sea-urchin
sperm \cite{GRAY1955}, so that the phase difference between the
first and the last particles of the tail is $4\pi$, and two waves
are present (see Fig.~\ref{spermmodel}). $f_s$ is the beating
frequency of the $s$-th sperm. The constant $c_{0,tail}$ determines
the average spontaneous curvature of the tail. $\varphi_s$ is the
initial phase of the first tail particle on the $s$-th sperm, and
$A$ is a constant related to the beating amplitude. We choose
$A=0.2$, which induces a beating amplitude $A_{tail}=3.2a$ of the
tail. As $t$ increases, a wave propagates along the tail, pushing
the fluid backward at the same time propelling the sperm forward. We
keep $A$, $k$, $T_s$, and $\varphi_s$ constant for each sperm during
a simulation. Although the spontaneous local curvature is
prescribed, the tail is elastic and its configuration is affected by
the viscous medium and the flow field generated by the motion of
neighboring sperm.

In order to avoid intersections or overlaps of different sperm,
we employ a shifted, truncated Leonard-Jones potential
\begin{equation}
V(r)=\left\{\begin{array}{l l}
  4\epsilon\left[\left(\frac{\sigma}{r}\right)^{12}
     -\left(\frac{\sigma}{r}\right)^6\right]+\epsilon, &r<2^{1/6}\sigma\\
  0, &r\geq2^{1/6}\sigma
\end{array}\right.
\end{equation}
between particles belonging to different sperm,
where $r$ is the distance between two particles. Parameters
$\sigma=1$ and $\epsilon=13.75$ are chosen.

During the MPC streaming step, the equation of motion of the sperm
particles is integrated by a velocity-Verlet algorithm, with a
molecular-dynamics time step $\Delta t_s=5\times 10^{-4}$, which
is $1/50$ of the MPC time step $\Delta t$.  The sperm only interacts
with the fluid during the MPC collision step. This is done by
sorting the sperm particles together with the fluid particles
into the collision cells and rotating their velocities relative to the
center-of-mass velocity of each cell.

Since energy is
injected into the system by the actively beating tails, we employ a
thermostat to keep the fluid temperature constant by rescaling all
fluid-particle velocities in a collision box relative to its
center-of-mass velocity after each collision step.
This procedure has the advantage that
the energy consumption per unit time of the sperm can be
easily extracted through the rescaling of the particle velocities.

We start with a single-sperm system with $c_{0,tail}=0$ and
$f=1/120$. With the other parameters given in the previous section,
our sperm model swims smoothly forward with the velocity
$u_{single}=0.016\pm0.001$.  Because of its large size, the
diffusion coefficient of a sperm due to the thermal fluctuations of
the MPC fluid is very small, on the order of $10^{-4}$
\cite{Mussawisade}. This implies that the time the sperm needs to
cover a distance of half the length of its flagellum by passive
diffusion is more than a factor $10^4$ larger than the time to
travel the same distance by active swimming. Therefore, diffusion
plays a negligible role in our simulations. The energy consumption
per unit time $P_{single}=25.2\pm2.4$
--- corresponding to $P_{single} \simeq 3000k_BT f$. Thus, we
estimate a Reynolds number $Re=2A_{tail}u_{single}/\nu \simeq 0.03$
for our sperm model, where $A_{tail}=3.2a$ is the beating amplitude
of the tail.

\section{Two-Sperm Simulations}
\label{sec:twosperm}

\subsection{Symmetric Sperm}
\label{sec:twosperm_symm}

Two sperm, S1 and S2, are placed inside the fluid, initially with straight
and parallel tails at a distance $d=5$ ({\em i.e.} with touching heads).
They start to beat at $t=0$ with different phases $\varphi_1$ and $\varphi_2$.
The initial positions of sperm do not matter too
much, because two freely swimming sperm always have the chance to
come close to each other after a sufficiently long simulation time.
We consider two sperm with the same beat frequency $f=1/120$,
and the same spontaneous curvature $c_{0,tail}=0$.

\begin{figure}
\subfigure[]{\resizebox{5cm}{!}{\includegraphics{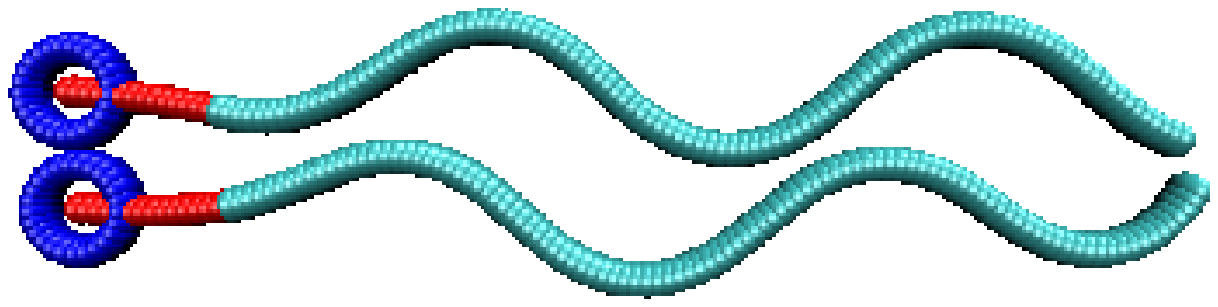}}}
\subfigure[]{\resizebox{5cm}{!}{\includegraphics{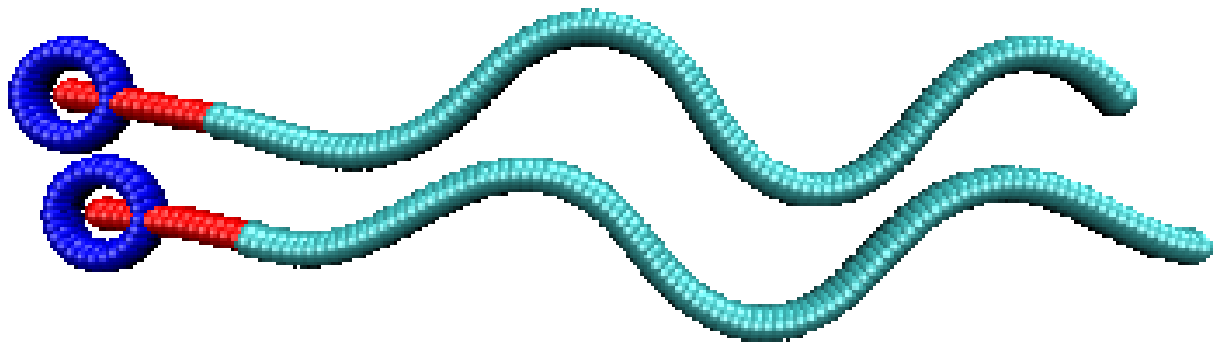}}}
\subfigure[]{\resizebox{5cm}{!}{\includegraphics{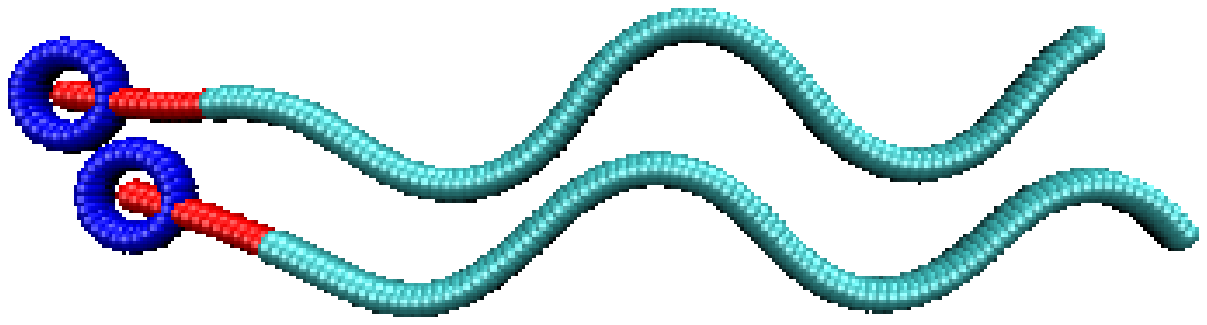}}}
\subfigure[]{\resizebox{5cm}{!}{\includegraphics{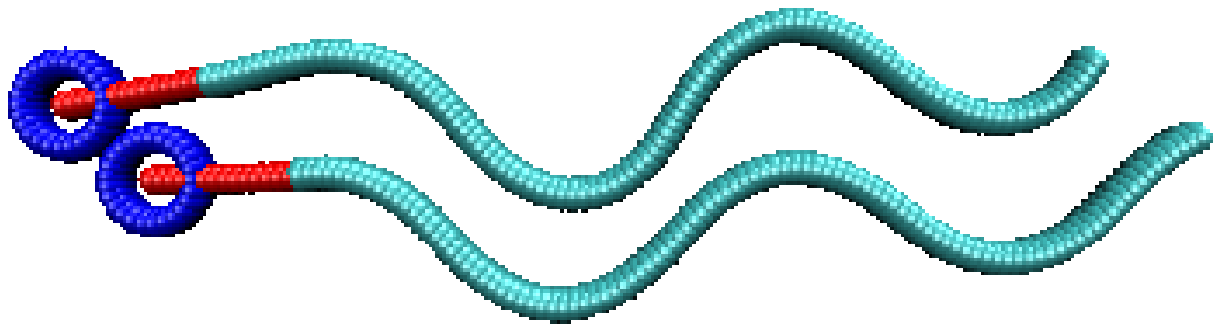}}}
\subfigure[]{\resizebox{5cm}{!}{\includegraphics{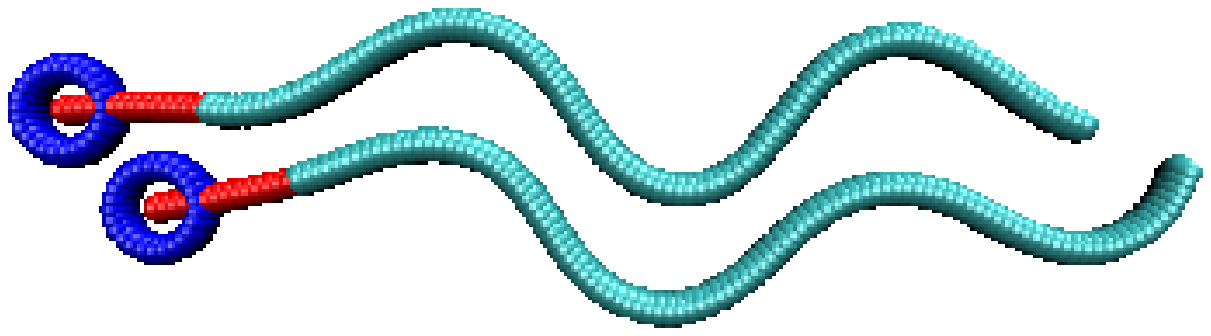}}}
\subfigure[]{\resizebox{5cm}{!}{\includegraphics{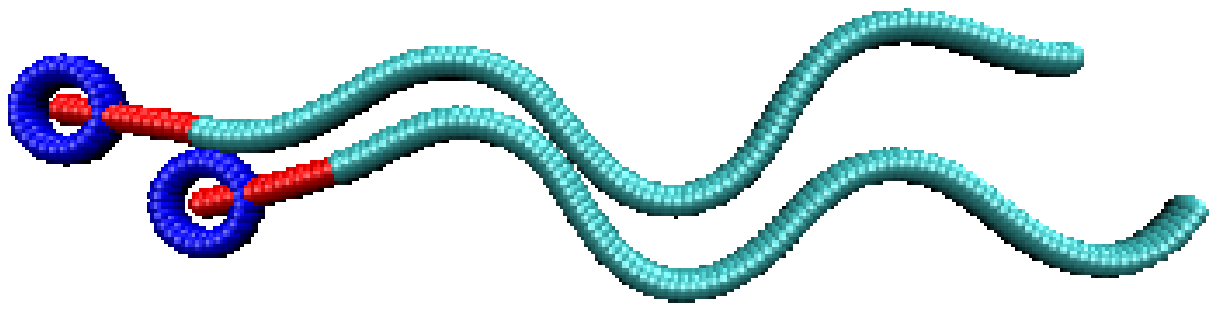}}}
\subfigure[]{\resizebox{5cm}{!}{\includegraphics{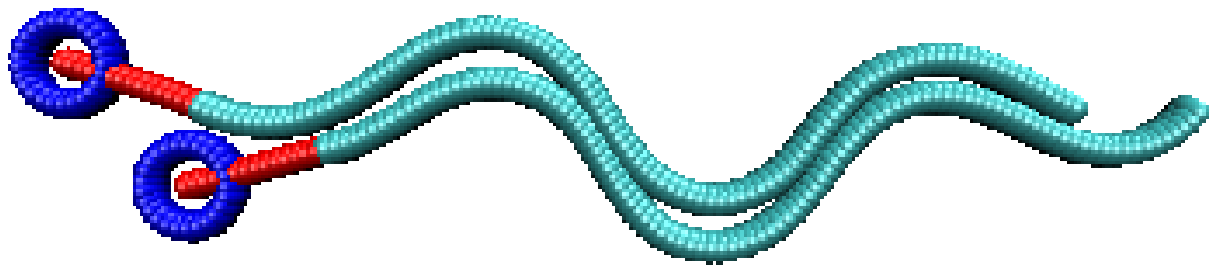}}}
\caption{(Color online) Snapshots of two sperm with phases
$\varphi_1$ (upper), $\varphi_2$ (lower), and phase difference
$\Delta\varphi=\varphi_2-\varphi_1=0.5\pi$. (a) $tf=1/6$ initial
position; (b) $tf=2/3$; (c) $tf=7/6$; (d) $tf=5/3$; (e) $tf=13/6$;
(f) $tf=10\frac{1}{6}$; (g) $tf=100\frac{1}{6}$. From (a) to (e),
the synchronization process takes place. The tails are already
beating in phase in (e). From (e) to (g), two synchronized sperm
form a tight cluster due to hydrodynamic attraction.}
\label{snapshot}
\end{figure}

In the dynamical behavior
of these hydrodynamically interacting sperm, two effects can be
distinguished, a short time ``synchronization" and a longer time
``attraction" process.
If the initial phase difference $\Delta\varphi=\varphi_2-\varphi_1$
at time $t=0$ is not too large, an interesting effect denoted
``synchronization" takes place, which is accomplished within a few
beats. This process is illustrated in
Fig.~\ref{snapshot}a-e by snapshots at different simulation times.
The synchronization time depends on the phase difference, and varies
from about two beats for $\Delta\varphi=0.5\pi$ (see Fig.~\ref{snapshot})
to about five beats for $\Delta\varphi=\pi$.
A difference in swimming velocities adjusts the relative positions of
the sperm. After a rapid transition, the velocities of two cells
become identical once their flagella beat in phase. Because the
initial distance between tails $d=5$ is smaller than the beating
amplitude $2A_{tail}=6.4a$, the sperm tails can touch when they start
to beat for $0.6\pi<\Delta\varphi<1.4\pi$. This geometrical effect is
reduced by the hydrodynamic interaction, which affects the beating
amplitude. In case contact occurs,  it accelerates
the synchronization. In order to avoid this direct interaction due to
volume exclusion, we have also performed simulations of two sperm with
initial distance $d=10$, and find the synchronized state achieved within
several beats, as in the simulations with $d=5$.
Thus, the synchronization effect is of purely hydrodynamic origin. Since
the beating phase at time $t$ is determined by $f$ and $\varphi_s$,
which are kept constant in our simulations, our model sperm can only
achieve synchronization by adjusting the relative position.

Our results are in good agreement with the prediction of
Taylor \cite{Taylor}, based on an analytical analysis of two-dimensional
hydrodynamics, that the viscous
stress between sinusoidally beating tails tends to force the two
waves into phase. The same phenomenon has also been observed by
Fauci and McDonald \cite{Fauci} in their simulations of sperm in the
presence of boundaries, and has been called ``phase-locking" effect.
A similar effect of undulating filaments immersed in a
two-dimensional fluid at low Reynolds number was seen by Fauci in
Ref.~\cite{Fauci2}.

Synchronization is a fast process, which is achieved in at most ten
beats in our simulations. Another hydrodynamic effect, which we denote
``attraction", takes much longer time. Two synchronized and separated
sperm gradually approach each other when they are swimming together,
as if there was some effective attractive interaction between them.
The only way in which the sperm can attract each other in our
simulations is through the hydrodynamics of the solvent.
This effect takes several ten beats to overcome the initial distance
of $d=5$ between the tails. The final state of attraction, in which the
sperm tails are touching tightly, is shown in Fig.~\ref{snapshot}g.

Fauci and McDonald \cite{Fauci}, did not see the hydrodynamic
attraction, because they considered a sperm pair confined
between two walls. As explained in Ref.~\cite{Fauci}, there is an
evident tendency for a single sperm to approach the wall. When two
parallel sperm are placed between the walls, there seems to be a
critical initial distance between the sperm, below which
synchronization occurs, and above which swimming towards the wall
occurs. Our understanding is that, in their simulations, the viscous
drag towards the walls was competing with the viscous attractive
effect between sperm. Hence in some cases, they could only see a
synchronization effect, and neither a clear towards-wall tendency
nor a distinguishable attraction effect. The hydrodynamic
attraction was masked by the presence of the walls.

\begin{figure}
\includegraphics[width=9cm]{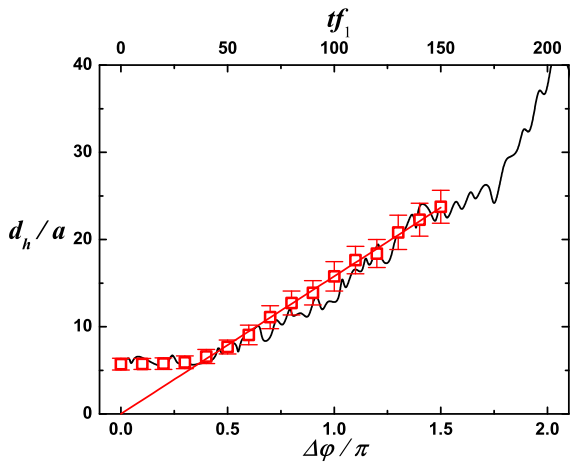}
\caption{(Color online) Head-head distance $d_h$ of two cooperating
sperm. Simulation data are shown for fixed phase difference (red,
$\square$), with error bars denoting the standard deviation. The
interpolating (red) line is a linear fit for  $0.4\pi <
\Delta\varphi <1.5\pi$. The distance $d_h$ is also shown as a
function of time $t$ (top axis) in a simulation with a $0.5\%$
difference in the beat frequencies of the two sperm (solid line). }
\label{symheaddis}
\end{figure}

To analyze the cooperating sperm pair in more detail, we choose the
head-head distance $d_h$ to characterize the attraction and
synchronization, because it is easy experimentally to track the head
position. The dependence of $d_h$ on the phase difference is
symmetric with respect to $\Delta\varphi=0$ because of the symmetry
of the sperm structure. Thus, we show in Fig.~\ref{symheaddis} only
results for $\Delta\varphi>0$.  There is a plateau at about $d_h=5a$
for $\Delta\varphi<0.4\pi$, which corresponds to the sperm heads
touching each other. For $\Delta\varphi>0.4\pi$, $d_h$ increases
linearly with $\Delta\varphi$. Finally, for $\Delta\varphi>1.5\pi$,
the phase difference is so large that the attraction is not strong
enough to overcome the thermal fluctuations and pull the sperm close
together. Although synchronization still occurs at the beginning,
the two sperm leave each other soon after.

Riedel et al.~\cite{Riedel} also see such a linear relation in their
experiments of sea-urchin sperm vortices. They define the beating
phase of a sperm by its head oscillation, and an angular position of
the sperm head within the vortex. In this way, the beating phase
difference of the sperm in the same vortex was found to have a linear
relation with the angular position difference, which corresponds to
the head-head distance in our simulations.

So far, we have considered sperm with a single beat frequency.
In nature, sperm of the same species always have a wide distribution
of beat frequencies. For example, the beat frequency of sea-urchin
sperm ranges from 30Hz to 80Hz \cite{GRAY1955}, and the frequency of
bull sperm ranges from 20Hz to 30Hz \cite{Riedel-Kruse07}. Thus, we
assign different beat frequencies to
two sperm, $f_1=1/120$ and $f_2=1/119.4$, corresponding to $\Delta
f/f_1\approx 0.5\%$, but set the same initial phases $\varphi_s=0$.
This implies that the phase difference of the beats between the two
sperm increases linearly in time,
\begin{equation}
    \Delta\varphi=2\pi(f_2-f_1)t \ .
\end{equation}
Fig.~\ref{symheaddis} shows the head-head distance versus time. It
agrees very well with the data for fixed phase differences. At
$tf_1=150$ where $\Delta\varphi \simeq 1.5\pi$, the sperm trajectories begin
to depart.

\begin{figure}
\includegraphics[width=9.0cm]{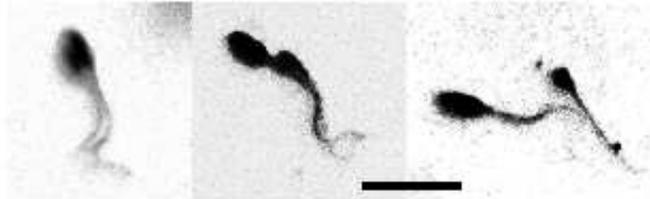}
\caption{Snapshots of two synchronized human sperm in experiment at
different times \cite{Kaupp08,humanexp}. (Left) Two sperm with
initially well synchronized tails and very small phase difference;
(Middle) the sperm are still swimming together and are well
synchronized after 4 seconds; a phase difference has developed;
(Right) the sperm begin to depart after 7 seconds. The scale bar
corresponds to a length of 25 $\mu$m.} \label{exphuman}
\end{figure}

Fig.~\ref{exphuman} shows two cooperating human sperm swimming in
an in-vitro experiment near a glass substrate \cite{Kaupp08,humanexp}. The
two sperm swim together for more than 6 seconds at
a beat frequency of approximately 8Hz. Their tails remain
synchronized during this time, while the head-head distance and
phase difference increases with time (see Fig.~\ref{exphuman}).
After a while, the sperm leave each other because the phase
difference becomes too large. There is no indication of a
direct adhesive interaction between the sperm.

\begin{figure}
\includegraphics[width=9cm]{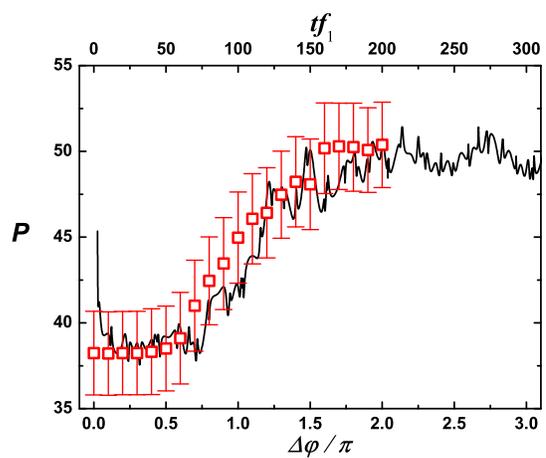}
\caption{(Color online) Energy consumption per unit time, $P$, of
two cooperating sperm. Symbols show simulation data for fixed phase
difference (red, $\square$), where error bars denote the standard
deviation. $P$ versus time $t$ in a simulation with a $0.5\%$
difference in the frequencies of two sperm (solid black line). }
\label{sympower}
\end{figure}

An interesting question is whether the cooperation of a sperm pair
reduces the energy consumption. Fig.~\ref{sympower} displays the
energy consumption of two sperm with the same beat frequency
$f=1/120$ as a function of the phase difference. The energy
consumption $P$ is nearly constant at small phase difference. It
increases for $\Delta\varphi\geq0.5\pi$ roughly linearly until it
reaches another plateau for $\Delta\varphi\geq1.5\pi$. The second
plateau corresponds to two sperm swim separately, so that energy
consumption is twice the value of a single sperm. Our results
are in agreement with the conclusion of Taylor \cite{Taylor} that
less energy is dissipated in the fluid if the tails are synchronized.

Fig.~\ref{sympower} also shows the energy consumption of two sperm
with $f_1=1/120$, $f_2=1/119.4$ and $\varphi_1=\varphi_2=0$ as a
function of time $t$. In this simulation, we start with two sperm
which are parallel and at a distance $d=5$. For $tf_1<25$, the energy
consumption decreases as the sperm are approaching each other. The
data agrees quantitatively very well with results for constant
$\Delta\varphi$, and reaches a plateau when the cooperating sperm
pair departs.

The synchronization and attraction also exists in our simulation of
swimming flagella without heads. In this case,
the time-reversal symmetry of Stokes flow implies that
no synchronization nor attraction is possible at zero Reynolds number.
In our simulations, the thermal
fluctuations and a finite Reynolds number break the time-reversal
symmetry.

\subsection{Asymmetric Sperm}

In nature, sperm have an abundance of different shapes.
In particular,
these shapes are typically not perfectly symmetric. The asymmetric shape
can cause a curvature of the sperm trajectory \cite{Jens,Elgeti2008}. For
example, sea-urchin sperm uses the spontaneous curvature of the tail to
actively regulate the sperm trajectory for chemotaxis
\cite{kaupp06,friedrich07}.  In our simulations, we
impose an asymmetry of the tail by employing a non-zero spontaneous
curvature $c_{0,tail}$.

\begin{figure}
\includegraphics[width=9cm]{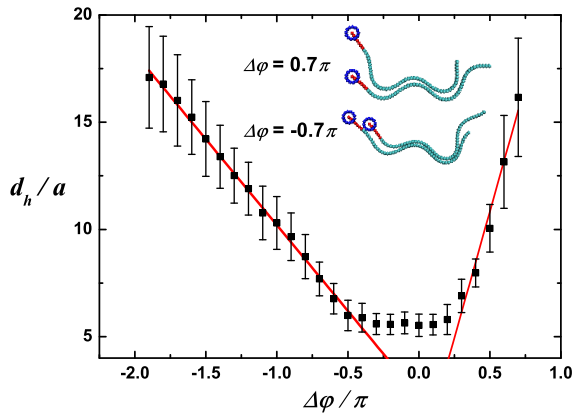}
\caption{(Color online) Head-head distance $d_h$ of two cooperating
sperm with spontaneous curvature $c_{0,tail}=0.04/a$ as a function
of the phase difference $\Delta\varphi$. The error bars represent
standard deviations. Lines are linear fits to the data in the range
$-1.9\pi<\Delta\varphi<-0.5\pi$ and $0.3\pi<\Delta\varphi<0.7\pi$,
respectively. The inset shows two typical conformations with
positive and negative phase difference.   } \label{bendheaddis}
\end{figure}

\begin{figure}
\includegraphics[width=9cm]{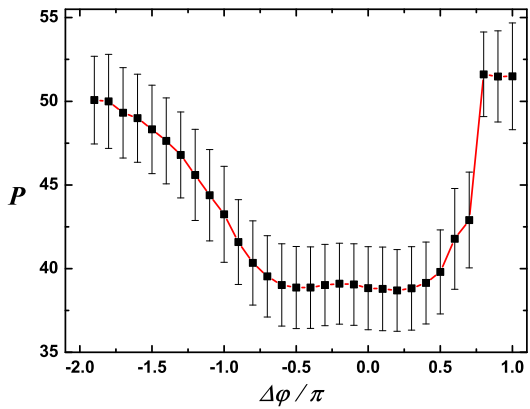}
\caption{(Color online) Energy consumption per unit time, $P$,
versus phase difference $\Delta\varphi$ of two sperm with
spontaneous curvature $c_{0,tail}=0.04/a$ of their tails. The error
bars represent standard deviations. } \label{bendpower}
\end{figure}

We consider curved sperm tails, with $c_{0,tail}=0.04/a$, which
results in a mean curvature of the trajectory of a single sperm of
$c_ta=0.041\pm0.009$. For sperm with curved tails, the head-head
distance $d_h(\Delta\varphi)$ is not symmetric about
$\Delta\varphi=0$, as shown in Fig.~\ref{bendheaddis}. Here
$\Delta\varphi$ is defined as the phase of the sperm on the inner
circle minus the phase of the sperm on the outer circle.
The steric repulsion of the heads causes a plateau of the head-head
distance at $d_h=5$ for small phase differences $\Delta\varphi$,
as in Fig.~\ref{symheaddis} for symmetric sperm.
For $\Delta\varphi<-\pi/2$ and $\Delta\varphi > \pi/4$, the
head distance increases linearly with increasing phase difference,
with a substantial difference of the slopes for $\Delta\varphi<0$ and
$\Delta\varphi>0$, see Fig.~\ref{bendheaddis}. The two sperm depart
when $\Delta\varphi>0.7\pi$.  For $\Delta\varphi\lesssim -2.0\pi$, the
sperm pair briefly looses synchronicity, but then rejoins with a
new phase difference $\Delta\varphi'=\Delta\varphi+2\pi$.

The energy consumption $P$ for sperm with spontaneous curvature
(see Fig.~\ref{bendpower}) also increases sharply at
$\Delta\varphi=0.8\pi$ and stays at the plateau with $P=51.0\pm2.8$
for larger $\Delta\varphi$, as the sperm are swimming separately.
However, for $\Delta\varphi<-\pi/2$, $P$ increases rather smoothly
until the cooperation is lost for large phase differences.

We conclude that the strong curvature of the tail breaks the
symmetry of the head-head distance $d_h$ and the energy consumption
$P$ in $\Delta\varphi$, but the effect of synchronization and
attraction are still present and play an important role in the
cooperation of sperm pairs.

\section{Multi-Sperm Systems}\label{multisperm}

When two sperm with the same beating period happen to get close and
parallel, they interact strongly through hydrodynamics and swim
together. With this knowledge of hydrodynamic interaction between
two sperm, we now study a system of 50 sperm in a simulation box of
$200\times200$ collision boxes, which corresponds to a density of
about three sperm per squared sperm length. The initial
position and orientation for each sperm are chosen randomly.
Considering that in real biological systems the beat frequency is
not necessary the same for all sperm, we perform simulations with
Gaussian-distributed beating frequencies. The initial phases of all
sperm are $\varphi_s=0$.

\begin{figure}
\subfigure[]{\resizebox{!}{5cm}{\includegraphics{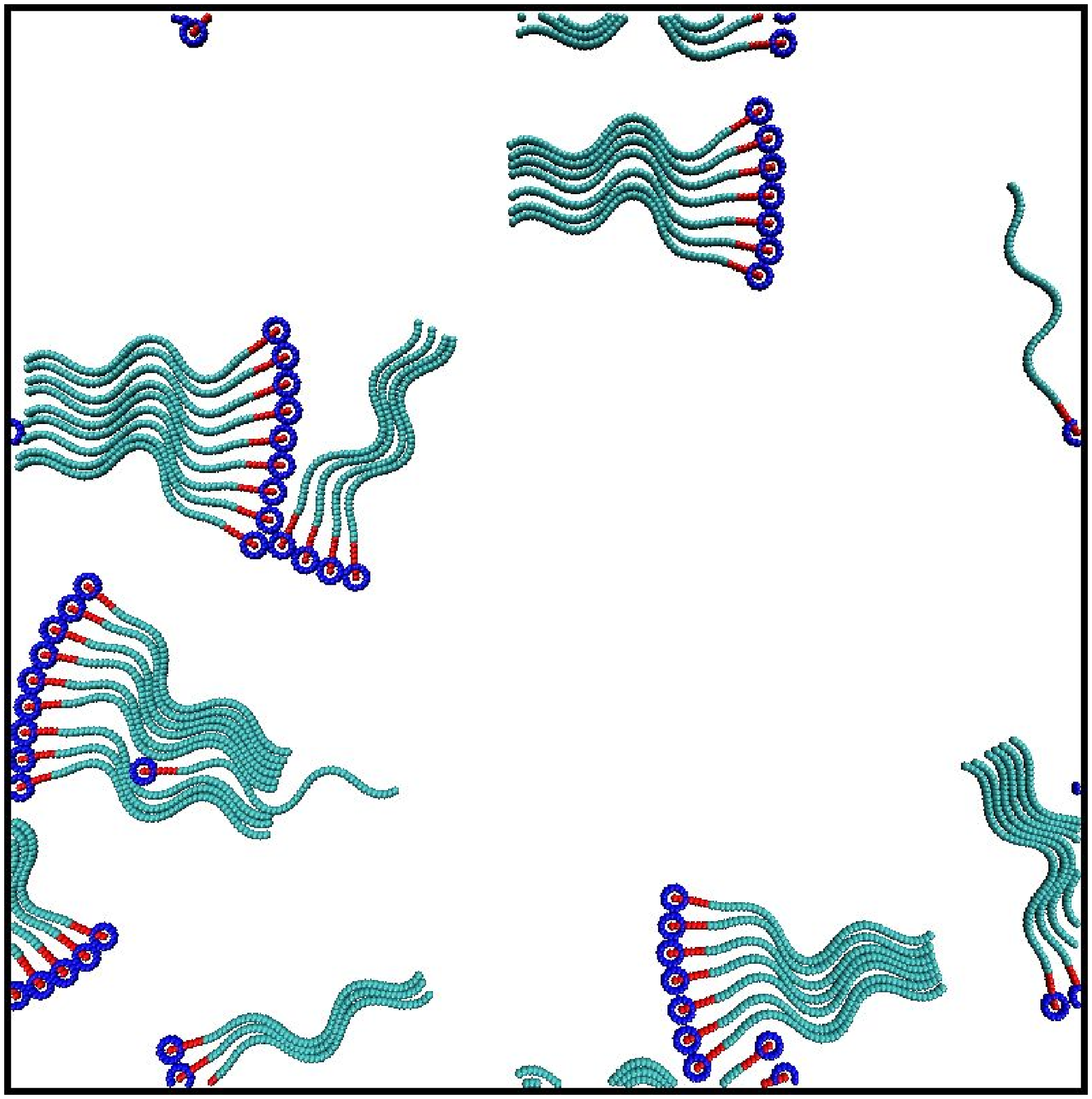}}}
\subfigure[]{\resizebox{!}{5cm}{\includegraphics{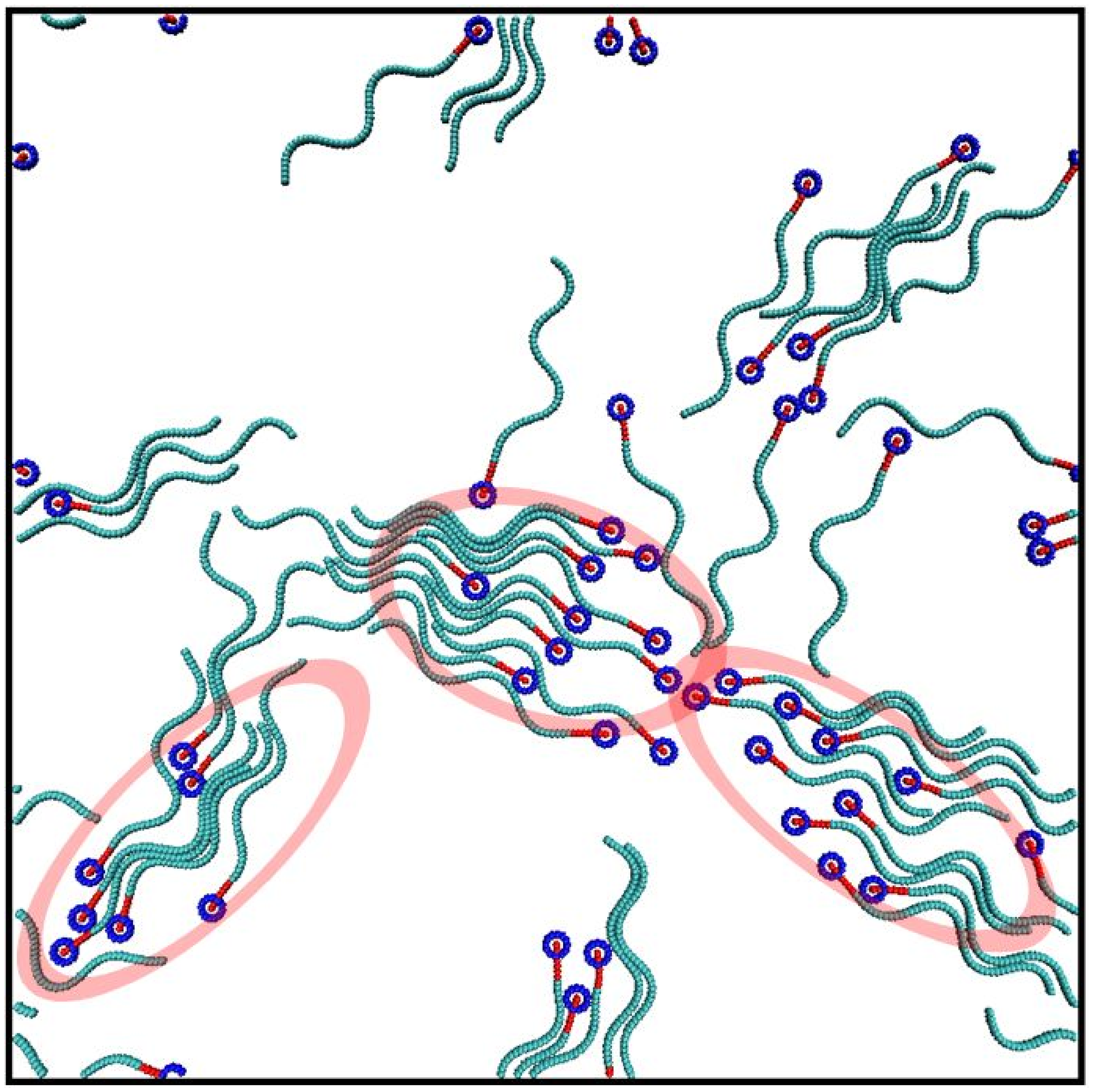}}}
\subfigure[]{\resizebox{!}{5cm}{\includegraphics{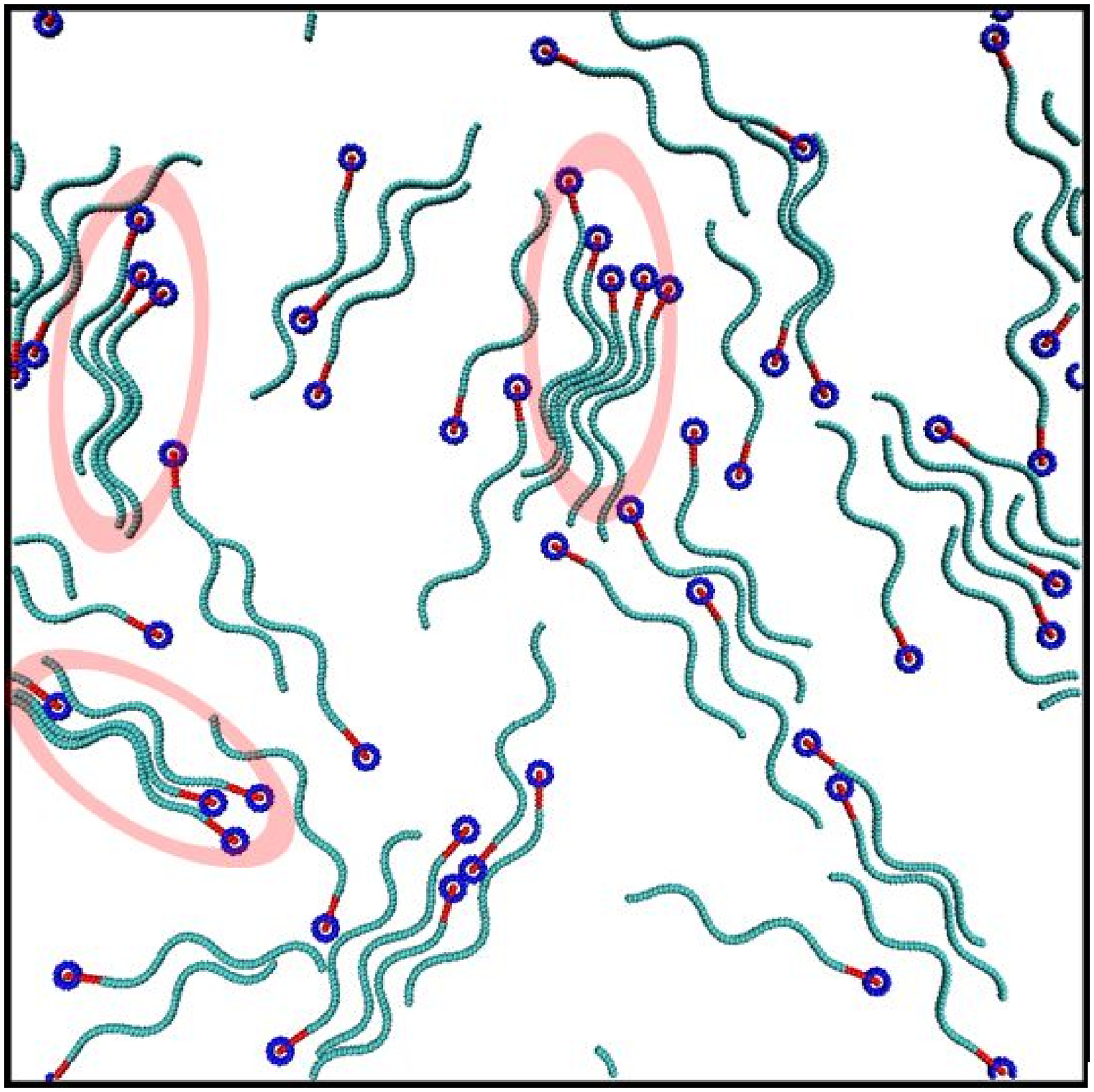}}}
\caption{(Color online) Snapshots from simulations of 50 symmetric
sperm with different widths $\delta_f$ of a Gaussian distribution of
beating frequencies. (a) $\delta_f=0$; (b) $\delta_f=0.9\%$; (c)
$\delta_f=4.5\%$. The red ellipses in (b) and (c) indicate large
sperm clusters. The black frames show the simulation boxes. Note
that we employ periodic boundary conditions.} \label{snapshotsym}
\end{figure}

We consider a system of symmetric sperm. Fig.~\ref{snapshotsym}
shows some snapshots of systems with different width
$\delta_f=\langle (\Delta f)^2 \rangle^{1/2}/\langle f \rangle$ of the
Gaussian frequency distribution. Here, $\langle (\Delta f)^2 \rangle$ is
the mean square deviation of the frequency distribution, and
$\langle f \rangle=1/120$ is the average frequency.

For $\delta_f=0$, once a cluster has formed, it does not
disintegrate without a strong external force. A possible way of
break-up is by bumping head-on into another cluster.
For $\delta_f>0$, however, sperm cells can leave a cluster after
sufficiently long time, since the phase difference to other cells in
the cluster increases in time due to the different beat frequencies
(compare Sec.~\ref{sec:twosperm_symm}). At the
same time, the cluster size can grow by collecting nearby free sperm or
by merging with other clusters. Thus, there is a balance between
cluster formation and break-up, as shown in the accompanying
movie \cite{movie}. Obviously, the average cluster size
is smaller for large $\delta_f$ than for small $\delta_f$ (see
Fig.~\ref{snapshotsym}).

\begin{figure}
\includegraphics[width=9cm]{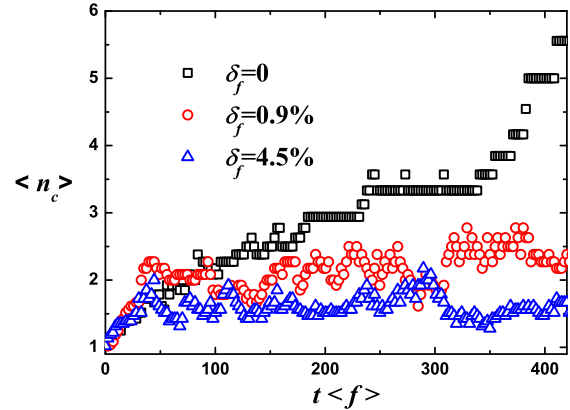}
\caption{(Color online) Time dependence of the average cluster size,
$<n_c>$, in a system of 50 symmetric sperm with various widths
$\delta_f$ of the frequency distribution, as indicated. }
\label{symcluster}
\end{figure}

\begin{figure}
\includegraphics[width=9cm]{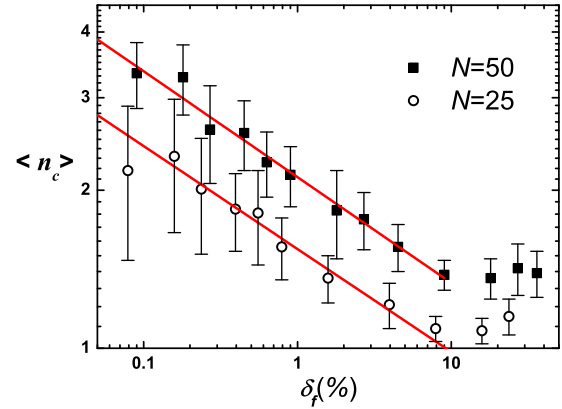}
\caption{(Color online) Dependence of the average stationary cluster
size, $<n_c>$, on the width of the frequency distribution
$\delta_f$. Data are shown for a 50-sperm system ($\blacksquare$)
and a 25-sperm system ($\circ$).  The lines indicate the power-law
decays $<n_c>=2.12\,\delta_f^{-0.201}$ (upper) and
$<n_c>=1.55\,\delta_f^{-0.196}$ (lower). } \label{symclusterave}
\end{figure}

\begin{figure}
\includegraphics[width=9cm]{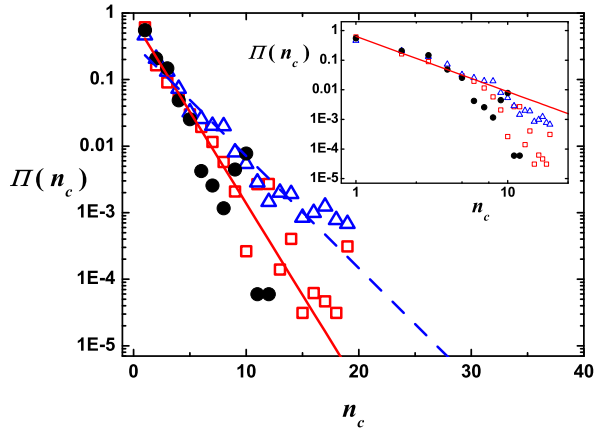}
\caption{(Color online) Cluster size distribution, $\Pi(n_c)$. Data
are shown for 25 sperm in a $200\times 200 a^2$ box ($\bullet$), 100
sperm in a $400 \times 400 a^2$ box ($\square$) [note that both
systems have the same sperm density], and 50 sperm in a $200 \times
200 a^2$ box ($\triangle$). The lines correspond to an exponential
distribution. The inset shows the same data in a double-logarithmic
representation. The line indicates a power law $n_c^{-1.8}$. }
\label{symclusterdis}
\end{figure}

To analyze the multi-sperm systems, we define a cluster as follows. If
the angle between vectors from the last to the first bead of the
tails of two sperm is smaller than $\pi/6$, and at the same time the
nearest distance between the tails is smaller than $4a$, which is
approximately 1/10 of the length of the tail, then we consider these
two sperm to be in the same cluster. By this definition, we find the
evolution of the average cluster size $<n_c>$ shown in
Fig.~\ref{symcluster}. Here, $<n_c>$ is the average number of sperm
in a cluster,
\begin{equation}
<n_c> = \sum_{n_c} n_c \Pi(n_c) \ ,
\end{equation}
where $\Pi(n_c)$ is the (normalized) cluster-size distribution. For
$\delta_f=0$, the average cluster size continues to increase with
time. Both systems in Fig.~\ref{symcluster} with $\delta_f>0$ reach
a stationary cluster size after about 50 beats.  The stationary
cluster size is plotted in Fig.~\ref{symclusterave} as a function of
the width $\delta_f$ of the frequency distribution. We find a decay
with a power law,
\begin{align}
<n_c>\sim\delta_{f}^{-\gamma}
\end{align}
with $\gamma=0.20\pm0.01$. The error for $\gamma$ is estimated
from a fit of the data for both 50-sperm and 25-sperm systems. The negative
power law indicates that the cluster size diverges when $\delta_f \to 0$.
This tendency is also implied by the continuously
increasing cluster size for $\delta_f=0$ in Fig.~\ref{symcluster}.
The cluster-size distribution in the stationary state
is shown in Fig.~\ref{symclusterdis}. Cluster-size distributions
have been studied in much simpler systems of self-propelled
particles, such as point particles with a constant magnitude
of their velocities, which adjust their traveling direction to
the direction of their neighbors \cite{Vicsek95}. In such simplified
models, a power-law
decay of the cluster-size distribution with an exponent in the
range of 1.5 to 1.9 have been found \cite{Huepe04}, followed by
a rapid decay for large cluster sizes due to finite-size effects.
Similarly, in a system of
self-propelled rods with volume exclusion, a crossover from power-law
behavior at small cluster-sizes to a more rapid decay for large
cluster sizes has also been found \cite{Peruani06}.
A power-law decay of the cluster-size distribution is indeed
consistent with our results for smaller cluster sizes, as shown
in the inset of Fig.~\ref{symclusterdis}. The rather similar value
of the exponent with that of Ref.~\cite{Huepe04} is probably
fortuitous.
We attribute the exponential decay of the cluster-size distribution
for larger cluster size, which is apparent in
Fig.~\ref{symclusterdis}, to finite-size effects. Simulations of
larger system sizes are required to confirm this conclusion.

\begin{figure}
\subfigure{\includegraphics[width=9cm]{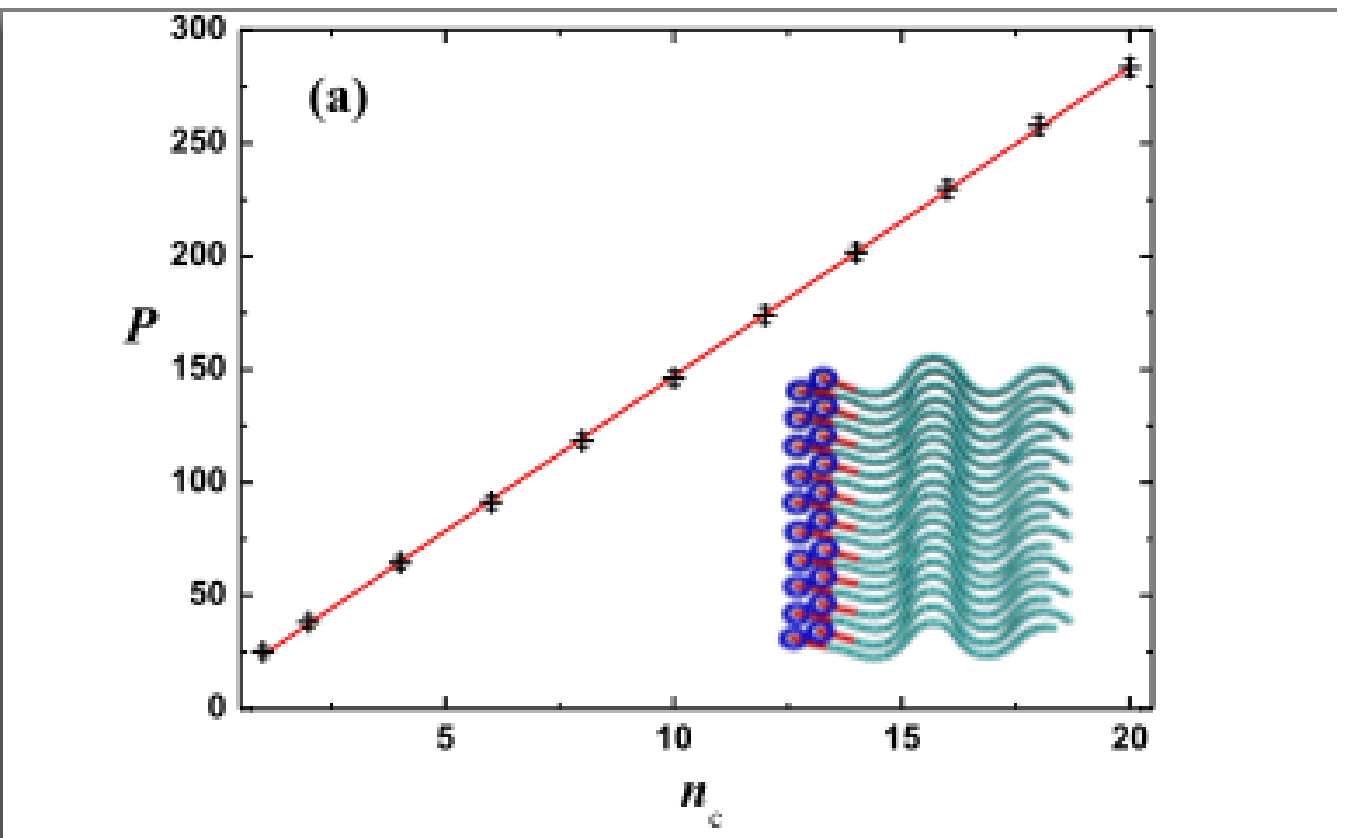}}
\subfigure{\includegraphics[width=9cm]{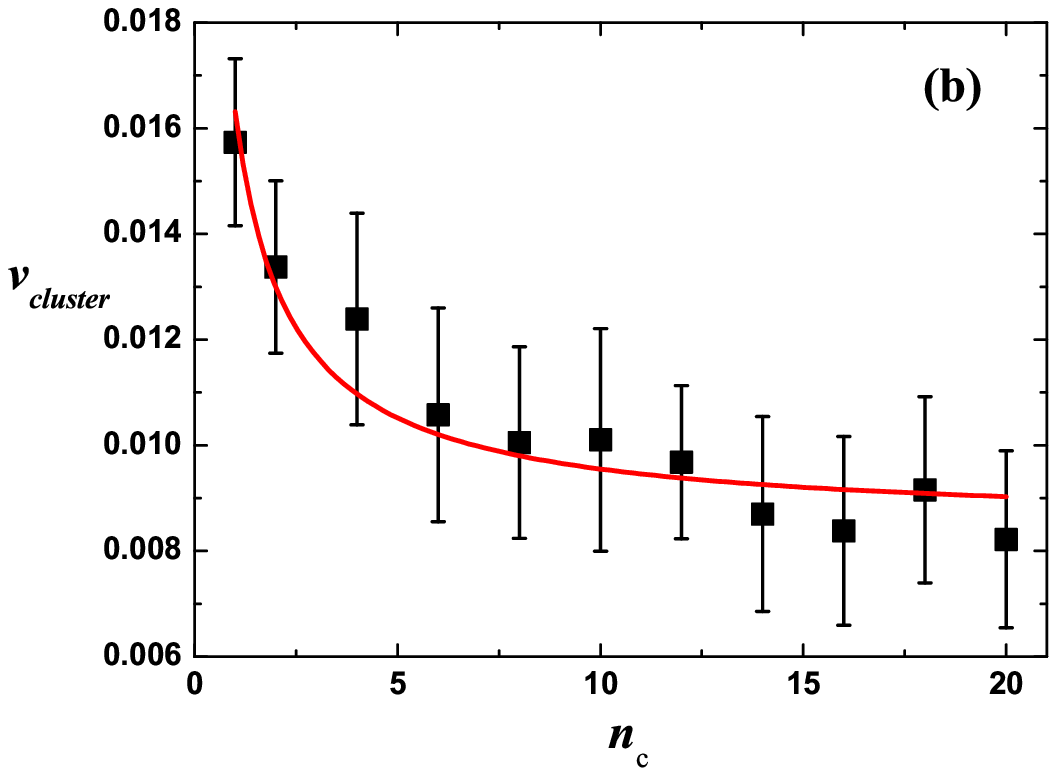}}
\caption{(Color online) (a) Energy consumption per unit time, $P$,
of sperm clusters as a function of cluster size $n_{c}$. Symbols
indicate simulations results. The fit line (red) is given by
$P=13.7n_c+10.1$. The inset shows an illustration of an arranged
cluster of 20 sperm. (b) Center of mass speed of sperm clusters as a
function of cluster size $n_c$. The fit line (red) is given by
$v=0.00334\,\sqrt{6.42+16.4/n_c}$.} \label{nspermpower}
\end{figure}

To analyze the energy consumption of sperm clusters, we consider a
special case where sperm of the same frequency are prearranged to
pack tightly and to be synchronized, as shown in
Fig.~\ref{nspermpower}a. A simple linear relationship between the
energy consumption of the sperm cluster and the cluster size is
shown in Fig.~\ref{nspermpower}a. From the linear fit of the data,
we obtain an energy consumption per sperm for an infinitely large
cluster, $P/n_c=13.7$. Thus, a freely swimming sperm can reduce its
energy consumption by almost a factor 2 by joining a cluster.

The swimming speed of a sperm cluster decreases slowly with increasing
sperm number, as shown in Fig.~\ref{nspermpower}b. When flagella
are very close, with distances smaller than the size of a MPC
collision box, hydrodynamic interactions are no longer properly
resolved. Instead, the collision procedure yields a sliding friction
for the relative motion of neighboring flagella.
Thus, the energy of the beat is not only used for propulsion, but
also to overcome the sliding friction. The energy consumption of
tail-tail friction is proportional to the number of neighbor pairs,
and the hydrodynamic resistance of moving the whole cluster is
proportional to the cluster size and speed.
Thus, the total energy consumption can be written as
\begin{equation}
\label{eq:speed}
P = C n_c v^2 + p_{f}(n_c-1) ,
\end{equation}
where $p_{f}$ is the energy consumption due to tail-tail friction,
and $C$ is a constant. With the relation $P=13.7\, n_c +10.1$
obtained above, the data for the cluster speed can be fitted to
Eq.~(\ref{eq:speed}), which yields $p_{f}=7.28$ and
$C=8.96\times10^4$. Thus, the cluster speed reaches a non-zero
asymptotic value $[(13.7-p_{f})/C]^{1/2}\simeq 0.0085$. for large
cluster size, about a factor 2 slower than a single sperm.

\section{Summary and Discussion}\label{conclusion}

We have simulated the hydrodynamic interaction between sperm in two
dimensions by the multi-particle collision dynamics (MPC) method. Two
effects of the hydrodynamic interaction were found in our
simulations. First, when two sperm are close in space and swimming
parallel, they synchronize their tail beats by adjusting their
relative position. This process can be accomplished in a very short
time, less than 10 beats. Second, two synchronized sperm have a
tendency to get close and form a tight pair. This process takes much
longer time then synchronization. It usually takes about 100 beats
to overcome a distance of $1/10$ tail length between sperm in our
simulations.

These hydrodynamic effects favor the cooperation of sperm in motile
clusters.
For a multi-sperm system, the average cluster size diverges if all
sperm have the same beating frequency. A distribution of frequencies
leads to a stationary cluster-size distribution with a finite average
cluster size, which decreases with a power law of the variance of the
frequency distribution. Furthermore, the average cluster size increases
with increasing sperm density.
The probability to find a cluster decreases
with a power law for small cluster sizes; an exponential decay
for large cluster sizes is attributed to finite-size effects.

In sperm experiments, large bundles have been found in some
species, like fish flies \cite{Hayashi,Hayashi2} and wood mouse
\cite{Moore,Moore2}. For fish-fly sperm, this has been attributed to
some agglutination
of the sperm heads to keep the size and structure of
the bundles.  Wood mouse sperm were released into an in-vitro
laboratory medium, initially in single cell suspension \cite{Moore}.
Within 10 minutes, large bundles containing hundreds or thousands of
sperm were formed as motile 'trains' of sperm. Motile bundles of
50-200 sperm were also found in the after-mating female's body, as
well as many non-motile single sperm. The hook structure on the head
of wood mouse sperm is believed to favor the formation of such huge
cluster in in-vitro experiments.

In our simulations, sperm clusters are always seen,
e.g. as marked in Fig.~\ref{snapshotsym}, after the
system has reached a dynamically balanced state of cluster sizes.
Thus, we predict that hydrodynamic synchronization and attraction
play an important role in the cluster formation of healthy
and motile sperm, such as the bundles and trains observed for fish-fly
and wood-mouse sperm at high concentrations, respectively. Furthermore,
since the cluster size decreases with increasing width $\delta_f$ of the
distribution of beat frequencies, our results are consistent with the
experimental observation that if the sperm are
hyperactivated \cite{Moore}, which is an abnormal beat mode,
or if some sperm are dead, the clusters fall apart.
\\

{\bf Acknowledgments:}
We thank U. Benjamin Kaupp, Luis Alvarez (CAESAR Bonn), and Luru Dai
(INB-1, Research Center J\"ulich) for stimulating discussions and
sharing of
their experimental data. YY acknowledges financial support by the
International Helmholtz Research School on Biophysics and Soft Matter
(``BioSoft").



\begin{thebibliography}{99}
\bibitem{humanexp}
The temperature in the human sperm experiment was about
$35-37\textordmasculine C$. The pictures were taken by fluorescence
microscopy. The sperm was loaded with Fluo4
(concentration 10 millimole/liter), a calcium indicator.
The measurement stage was moved to keep the swimming sperm in the
field of view.
\end{thebibliography}
\end{document}